\documentclass[showpacs,preprintnumbers,twocolumn]{revtex4-1}
\usepackage{graphicx}
\usepackage{amsmath}
\usepackage{amssymb}
\begin{document}

\title{Efficiency at maximum power for classical particle transport.}

\author{ Christian Van den Broeck}
\affiliation{
Hasselt University, B-3590 Diepenbeek, Belgium.}

\author {Katja Lindenberg }
 \affiliation{Department of Chemistry and Biochemistry and BioCircuits Institute, University
of California San Diego, 9500 Gilman Drive, La Jolla, CA 92093-0340, USA}


\begin{abstract}
We derive the explicit analytic expression for the efficiency at maximum power in a simple model of classical particle transport.
\end{abstract}
\maketitle
\section{Introduction}\label{intro}
Over the past few years, there has been considerable interest in the study of small scale systems, with special emphasis on the issue of efficiency  \cite{efficiency}. It was shown, on the basis of general thermodynamic arguments \cite{chris1} and using stochastic thermodynamics \cite{esp1}, that the efficiency at maximum power $\eta^\star$ of a thermal engine, operating between a hot and cold bath at temperatures $T^{(1)}$ and $T^{(2)}$, respectively, possesses universal properties when expanded in terms of the Carnot efficiency $\eta_C=1-T^{(2)}/T^{(1)}$:
\begin{equation}
\eta^\star=\frac{\eta_C}{2}+\frac{\eta_C^2}{8}+\ldots .
\end{equation}
This result is valid for ``strong coupling," meaning that the particle and energy fluxes are proportional to each other. The value $1/8$ for the coefficient of the quadratic term in addition requires a left/right symmetry, i.e., reversal of fluxes upon inversion of forces. The above universality ultimately derives from the reversibility of the underlying microscopic laws \cite{chris2}.
It has been verified in various  models \cite{efficiency}, including transport of electrons through a quantum dot \cite{esp2} and of photons in a maser model \cite{esp1}.
The purpose of this brief report is to present an analogous calculation for the transport of classical particles. As expected, universality is reproduced. An additional benefit is that, contrary to the case of quantum transport, an explicit analytic expression is obtained for $\eta^\star$, namely,
\begin{equation}\label{main}
\eta^\star=\frac{\eta_C^2}{\eta_C-(1-\eta_C)\ln(1-\eta_C)}=\frac{\eta_C}{2}+\frac{\eta_C^2}{8}+\ldots .
\end{equation}

\section{Model and Master equation}\label{mme}

We consider  a (small) reservoir of classical  non-interacting particles which for simplicity all have the same energy $\epsilon$. This small reservoir constitutes our system. The number of particles in the system will be denoted by $n$.   
This system exchanges particles  with several other  particle reservoirs ${\nu}$ with temperatures $T^{(\nu)}$ and chemical potentials $\mu^{(\nu)}$, respectively.
 We assume that the exchange can be described by  Markovian dynamics.  This is the case, for example, if the transitions between system and reservoirs correspond to  thermally activated processes over sufficiently high potential barriers. The probability distribution $p_n(t)$ for the system to be in state $n$ at time $t$ thus obeys the following master equation ($n \in \bold{N}$, quantities with negative $n$-index being zero by definition):   
\begin{eqnarray}\label{ME}
\dot{p}_n&=&W_{n,n-1}{p}_{n-1}+W_{n,n+1}{p}_{n+1}-(W_{n+1,n}+W_{n-1,n}){p}_{n},\nonumber\\
\end{eqnarray}
where $W_{n+1,n}$ and $W_{n-1,n}$ are the rates (probabilities per unit time) for transitions $n\rightarrow n+1$ (the system gains one particle) and  $n\rightarrow n-1$ (loses one particle), respectively.

From here on we are interested in the steady state operation of our system. The corresponding probability distribution ${p}^{st}$ is determined by the set of equations
\begin{eqnarray}\label{pst}
W_{n,n-1}{p}^{st}_{n-1}+W_{n,n+1}{p}^{st}_{n+1}-(W_{n+1,n}+W_{n-1,n}){p}^{st}_{n}=0,\nonumber\\
\end{eqnarray}
or (since the flux is zero at the boundaries $n=0$ and $n=\infty$),
 \begin{eqnarray}\label{pst2}
W_{n,n-1}{p}^{st}_{n-1}=W_{n-1,n}{p}^{st}_{n}.
\end{eqnarray}
Note that the one-step hopping dynamics of our master equation has the peculiarity that the above formal condition of detailed balance is satisfied, at least with respect to the total transition rates $W$. This however does not necessarily correspond to true equilibrium, which requires detailed balance at the level of each of the separate processes taking place in the system, as we will see below.

To proceed further, we have to specify the transition rates. We first assume that the exchanges of particles with different reservoirs are independent, so that the corresponding rates $W^{(\nu)}$ add up to the total rate $W$, that is,
\begin{eqnarray}
W_{n+1,n}=\sum_\nu W^{(\nu)}_{n+1,n},\\  
W_{n-1,n}=\sum_\nu W^{(\nu)}_{n-1,n}.
\end{eqnarray}
Second, statistical mechanics imposes physical constraints on the separate  rates $W^{(\nu)}$. Let us  suppose that the contact is broken with all reservoirs except for reservoir $\nu$. The stationary state should in this case reproduce the equilibrium state of the system in contact with this reservoir,  $p^{st}={p}^{eq,(\nu)}$. As is well know from equilibrium statistical mechanics, this is  the so-called grand canonical distribution, which for classical non-interacting (ideal) particles is a Poisson distribution \cite{wannier},
\begin{eqnarray}\label{eq1}
 {p}_{n}^{eq,(\nu)}=\frac{\{\bar{n}^{(\nu)}\}^n}{n!}e^{-\bar{n}^{(\nu)}}.
 \end{eqnarray}
The average particle occupation while in contact with reservoir $\nu$, $\bar{n}^{(\nu)}$, is given by
 \begin{eqnarray}\label{eq2}
 \bar{n}^{(\nu)}=e^{-x_{\nu}},
 \end{eqnarray}
where we have introduced the dimensionless quantity $x_{\nu}$,
\begin{eqnarray}
x_{\nu}=\beta^{(\nu)}({\epsilon-\mu^{(\nu)}}),
\end{eqnarray}
and $\beta^{(\nu)}=1/k_BT$ ($k_B=$ Boltzmann's constant).
One can rewrite this equation in the more familiar form
$\mu^{(\nu)}=\epsilon+k_B T^{(\nu)} \ln\bar{n}^{(\nu)}$.

The requirement that  $p^{st}={p}^{eq,(\nu)}$ for $W=W^{(\nu)}$ leads to the following  ``genuine" condition of detailed balance for these rates with respect to its equilibrium distribution:
\begin{eqnarray}\label{db}
W^{(\nu)}_{n,n-1}{p}_{n-1}^{eq,(\nu)}=W^{(\nu)}_{n-1,n}{p}_{n}^{eq,(\nu)}.
\end{eqnarray}
In the following we will adopt the standard choice of transition rates \cite{vankampen}   that satisfy this condition (cf. law of mass action), namely,
\begin{eqnarray}\label{db2}
W^{(\nu)}_{n+1,n}=k^{(\nu)}_{+}\nonumber\\
W^{(\nu)}_{n-1,n}=n\;k^{(\nu)}_{-},
\end{eqnarray}
with the $n$-independent rates obeying the balance condition
 \begin{eqnarray}\label{db3}
{k^{(\nu)}_+}={k^{(\nu)}_-}\bar{n}^{(\nu)}.
\end{eqnarray}

\section{Thermal Engine}
 
Having identified the thermodynamically correct expressions (\ref{db2}) and (\ref{db3}) for the transition probabilities, we can proceed to a stochastic thermodynamic analysis (see \cite{chris3} for a brief review) of  a system  coupled to several reservoirs, with different temperatures and chemical potentials.  One easily verifies that the steady state solution of (\ref{pst}), is again a Poisson distribution,
\begin{eqnarray}\label{eq3}
 {p}_{n}^{st}=\frac{\bar{n}^n}{n!}e^{-\bar{n}},
 \end{eqnarray}
with a steady state average number of particles reflecting the influence of each reservoir [compare with (\ref{db3})]:
\begin{eqnarray}\label{nst}
 \bar{n}=\frac{\sum_\nu k_+^{(\nu)}}{\sum_\nu k_-^{(\nu)}}.
 \end{eqnarray}
This is most easily demonstrated by showing that the generating function is given by
$\sum_n s^n p_n^{st}=e^{(s-1)\bar{n}}$

Even though  Poissonian, this distribution corresponds to a nonequilibrium  steady state.  Using    (\ref{db2})-(\ref{nst}), we find the following explicit expressions for the separate particle, energy,  and heat fluxes from each reservoir $\nu$ into the system:
\begin{eqnarray}\label{flu}
J_N^{(\nu)}&=&\sum_n (W^{(\nu)}_{n,n-1}{p}^{st}_{n-1}-W^{(\nu)}_{n-1,n}{p}^{st}_{n})\nonumber\\
&=&k_+^{(\nu)}-k_-^{(\nu)}\bar{n}\\
J_E^{(\nu)}&=&\epsilon J_N^{(\nu)}\\
J_Q^{(\nu)}&=&J_E^{(\nu)}-\mu^{(\nu)}J_N^{(\nu)}=(\epsilon-\mu^{(\nu)})J_N^{(\nu)}.
\end{eqnarray}
 Note that the fluxes from each reservoir to the system are strongly coupled, i.e.,  energy, heat  and particle flux  are proportional to each other. Furthermore, the fluxes from  reservoir $\nu$ are only zero when $\bar{n}=\bar{n}^{(\nu)}$, i.e. when the steady state distribution is the equilibrium distribution, cf. (\ref{eq1}) and (\ref{eq3}). 
The above Poissonian steady state (\ref{eq3}) does not, in general, obey detailed balance with respect to the separate rates $W^{({\nu})}$, and implies the following non-zero entropy production:
\begin{eqnarray}\label{ep}
\dot{S}_i&=&k_B \sum_{\nu}\sum_n (W^{(\nu)}_{n,n-1}{p}^{st}_{n-1}-W^{(\nu)}_{n-1,n}{p}^{st}_{n})\ln\frac{W^{(\nu)}_{n,n-1}{p}^{st}_{n-1}}{W^{(\nu)}_{n-1,n}{p}^{st}_{n}}\nonumber\\
&=&k_B \sum_{\nu}(k_+^{(\nu)}-k_-^{(\nu)}\bar{n})\ln\frac{k_+^{(\nu)}}{k_-^{(\nu)}\bar{n}}\nonumber\\
&=&\sum_{\nu}J_N^{(\nu)} X_N^{(\nu)}\geq 0,
\end{eqnarray}
where we have introduced the thermodynamic forces:
\begin{equation}
X_N^{(\nu)}= k_B \ln\frac{\bar{n}^{(\nu)}}{\bar{n}}.
\end{equation}

Of particular interest to us is the situation in which a heat current from a hot to a cold reservoir is used to drive particles uphill from low to high chemical potential. To investigate this case in more detail, we henceforth focus on the case of only two reservoirs $\nu=1,2$, with reservoir $1$ the hot reservoir and $2$ the cold one, $T^{(1)}\geq T^{(2)}$. 
Note that we have not taken the thermal energy of the particles into account as this would require the consideration of a third thermal reservoir, making the comparison with Carnot efficiency more involved. At  the steady state, one finds, using (\ref{eq2}) and (\ref{db3}), the following explicit results for the fluxes:
\begin{eqnarray}
J_N^{(1)}&=&-J_N^{(2)}
=\kappa(e^{-x_1}-e^{-x_2})\\
J_E^{(1)}&=&-J_E^{(2)}=\epsilon J_N^{(1)} \\
J_Q^{(1)}&=&k_B T^{(1)}x_1J_N^{(1)},
\end{eqnarray}
where we have introduced the rate
\begin{equation}
\kappa=\frac{k_-^{(1)}k_-^{(2)}}{k_-^{(1)}+k_-^{(2)}}.
\end{equation}

The entropy production reduces to the simple expression
\begin{eqnarray}\label{ep2}
\dot{S}_i=k_B \kappa (x_2-x_1)(e^{-x_1}-e^{-x_2})\geq 0.
\end{eqnarray}

The power $\cal{P}$ of the engine, being the amount of net chemical energy produced per unit time, is given by
\begin{eqnarray}
{\cal P}&=& (\mu_2-\mu_1) J^{(1)}_N\\
&=&\kappa\; k_B T^{(1)}\;[x_1-(1-\eta_c)x_2]\;(e^{-x_1}-e^{-x_2}).\nonumber
\end{eqnarray}
The corresponding efficiency reads:
\begin{eqnarray}\label{eff0}
\eta=\frac{(\mu_2-\mu_1)J^{(1)}_N}{J^{(1)}_Q}= 1-(1-\eta_c)\frac{x_{2}}{x_{1}}
\end{eqnarray}
Before turning to the main issue of efficiency at maximum power, we first note that equilibrium, i.e. zero entropy production, cf. (\ref{ep2}), is attained when $x_1=x_2$. This does not require that the temperatures $T^{(1)}$ and $T^{(2)}$and chemical potentials $\mu^{(1)},$ and $\mu^{(2)}$ be separately equal, a feature which is well-known for strongly coupled systems \cite{chris4}. In the vicinity of such a point, the machine can operate reversibly, and its efficiency attains Carnot efficiency, $\eta=\eta_C$, cf. (\ref{eff0}). 

Let us now turn our attention to the point of maximum power. From
\begin{eqnarray}
\frac{\partial{\cal P}}{\partial x_1}=\frac{\partial{\cal P}}{\partial x_2}=0
\end{eqnarray}
one finds:
\begin{eqnarray}
 x_1&=&1-(1-\eta_C)\frac{\ln(1-\eta_C)}{\eta_C}\\
 x_2&=&1-\frac{\ln(1-\eta_C)}{\eta_C}.\\
 \end{eqnarray}
The corresponding efficiency reads:
\begin{equation}\label{main2}
\eta^\star=\frac{\eta_C^2}{\eta_C-(1-\eta_C)\ln(1-\eta_C)}=\frac{\eta_C}{2}+\frac{\eta_C^2}{8}+...,
\end{equation}
which displays the expected universality announced earlier.

\section{Discussion}
We close with a number of additional comments.
The above result is identical to the one obtained for a model based on particle transport via Kramers' escape \cite{tu}. This can be understood from the fact that our model reduces  to this case when the outgoing rates $k_-$  become very large. In this limit the number of particles in the system goes to zero, and the only remaining processes are the thermally activated transitions from one reservoir into another, via fast passage through the system, which plays the role of a short-lived activated state.
The fact that our more general model reproduces the same result as Kramers' escape suggests that (\ref{main}) may have a wider applicability in classical transport.
The above results for particle flux, power, entropy production  and efficiency at maximum power are also reproduced by taking the classical limit in the problem  of electron transport through a quantum dot \cite{esp2}. 

The above model for particle transport can  be represented as a  simple chemical reaction, namely,
\begin{equation}
X \rightleftarrows X^{(\nu)} 
\end{equation}
where $n$ represents the number of particles of species $X$, which can transmute into the species $X^{(\nu)}$ whose chemical potential is fixed by reservoir constraints.  This representation is  most natural in an isothermal system, $T^{(\nu)}=T$. 

Finally, we mention a similarity of the model studied here with that of an underdamped Brownian particle in contact with several heat baths. The Langevin equation for such a particle has the following form:
\begin{equation}
m\dot{v}=\sum_{\nu}\{-\gamma^{(\nu)} v+\sqrt{\gamma^{(\nu)}T^{(\nu)}} \xi^{(\nu)}\},
\end{equation}
where $\xi^{(\nu)}$ are independent normal white noises. The noise intensity is chosen in accordance with the fluctuation-dissipation theorem so that the stationary distribution when in contact with each reservoir separately reduces to the corresponding Maxwellian velocity distribution.
As in the model studied here, the stationary distribution of the system in simultaneous contact with multiple reservoirs has an "equilibrium shape", i.e., a Maxwellian distribution, however at a temperature which is the geometric mean of the bath temperatures.  Detailed balance is broken at the level of the exchange between particle and separate reservoirs, with the Brownian particle functioning as a thermal contact between reservoirs. This Brownian model has been studied in great detail, revealing detailed properties of the corresponding nonequilibrium steady state \cite{bm}. The model presented here has the additional advantage of allowing both heat and particle transport. We therefore expect it to be  an interesting candidate for revealing further properties of  nonequilibrium steady states.

\section{Acknowledgments}
This research is supported
in part by the research network``Exploring the Physics of Small Devices" of the European Science Foundation and by the NSF under Grant No. PHY-0855471.


\end{document}